\documentclass[useAMS,usenatbib]{mnras}
\usepackage{amsmath}
\usepackage{amssymb}
\usepackage{graphicx}% Include figure files
\usepackage{dcolumn}% Align table columns on decimal point
\usepackage{bm}% bold math
\usepackage{ulem}

\title[Energy of Quantum Coulomb Liquid] 
{Energy of Quantum Coulomb Liquid}

\author[D.A. Baiko]
{D.A. Baiko\thanks{E-mail:baiko@astro.ioffe.ru} \\
Ioffe Institute, Politekhnicheskaya 26, 194021 Saint Petersburg, 
Russian Federation}

\begin{document}

\label{firstpage}
\date{Accepted; Received ; in original form}

\pagerange{\pageref{firstpage}--\pageref{lastpage}} \pubyear{2018}

\maketitle

\begin{abstract}
%\begin{linenumbers}
Using Metropolis method to compute path integrals, 
the energy of quantum strongly-coupled Coulomb liquid
($1 \leq \Gamma \leq 175$) composed of distinguishable atomic nuclei
and uniform incompressible electron background is calculated from
first principles. The range of considered temperatures and densities
represents fully-ionized layers of white dwarfs and neutron stars.
In particular, the results allow one to determine reliably the heat
capacity of ions in dense fluid stellar matter, which is a crucial 
ingredient for modelling thermal evolution of compact degenerate stars.
%\end{linenumbers}
\end{abstract}


\section{Introduction}
One-component plasma (OCP) is a system of identical point charges 
immersed into a uniform incompressible background of 
opposite charge to ensure overall charge-neutrality. This system is 
of paramount importance for astrophysics of degenerate stars, white 
dwarfs and neutron stars, where the point charges are bare atomic nuclei 
(ions) while the degenerate nearly ideal electron gas constitutes the 
background. For both types of stars, mass densities, at which matter 
can be viewed as the OCP, cannot be too low to ensure that atoms are 
fully ionized and electron screening is, to a good 
approximation, negligible. In white dwarfs, the OCP phase may extend 
all the way down to the center of the star (i.e. up to the highest 
available densities); however, several 
different types of atomic nuclei may be present at some densities 
simultaneously (e.g., C and O). Studying such multi-component plasma 
is beyond the scope of the present paper. In neutron stars, matter 
can be treated as the OCP at least in the outer crust, i.e. all the way 
up to the neutron-drip density, 
$\rho_{\rm d} \approx 4.3 \times 10^{11}$ g cm$^{-3}$ (with the same 
caveat regarding multi-ionic mixtures). It is possible 
that the same treatment remains valid up to higher, subnuclear, 
densities (in the inner crust), if the effect of dripped neutrons on 
the inter-ionic interaction is negligible.

Thermodynamic state of the OCP is described by just two 
dimensionless parameters. The Coulomb-coupling parameter 
$\Gamma = Z_i^2 e^2 /(aT)$ measures the strength of the typical 
potential energy with respect to the typical kinetic energy of ions. 
In this case $T$ is the temperature ($k_{\rm B} \equiv 1$), $Z_i$ is 
the ion charge number, $e$ is the 
elementary charge, and $a = (4 \pi n/3)^{-1/3}$ is the ion-sphere 
radius ($n$ is the ion number density). If one neglects the quantum 
aspect of the ions' motion, $\Gamma$ alone determines fully
thermodynamics of the OCP. In particular, $\Gamma=1$ signals the 
gas-liquid transition while $\Gamma=\Gamma_{\rm m} \approx 175$ 
corresponds to the liquid-solid transition. The strength of the ionic 
quantum effects is measured by the second parameter, which can be 
chosen as the ratio of $T$ to the ion plasma temperature 
$T_{\rm p} = \hbar \sqrt{4 \pi n Z_i^2 e^2 /M_i}$, where $M_i$ is the 
ion mass. In a quantum system, one has $T \ll T_{\rm p}$. Another often 
used parameter is the ratio 
$r_{\rm s} = a / a_0$, where $a_0 = \hbar^2 /(M_i Z_i^2 e^2)$ is the 
ionic Bohr radius. These quantities are related to each other as
\begin{equation}
     \frac{T}{T_{\rm p}} 
         = \frac{1}{\Gamma} \sqrt{\frac{r_{\rm s}}{3}}~.
\label{TTp}
\end{equation}

Thermodynamics of the OCP has been studied by many authors. 
In particular, the classic OCP has been studied since the 60's
\citep{BST66} 
by Monte Carlo (MC) and Molecular Dynamics (MD) methods and its
thermodynamic properties are now firmly established 
\citep[e.g.,][]{SDD82,DS99,C99}. 
The importance of the ionic quantum effects in white dwarfs (for 
cooling, in particular) has been 
appreciated since about the same time \citep[e.g.,][]{VH79} and was 
emphasized once again by \citet{CADW92}. However, there was no practical 
way to incorporate them. \citet{LVH75} have constructed
the first quantitative evolutionary model of white dwarfs with realistic
thermodynamics but the treatment of the quantum effects
was incomplete. 

An important study of the quantum OCP from first principles 
has been conducted by \citet{JC96}. These authors used Path Integral 
Monte Carlo (PIMC) with 54 distinguishable particles in a simulation 
cell with periodic boundary conditions. For the liquid phase, they 
have calculated 
the thermal energy at $\Gamma \leq 160$ and $r_{\rm s} = 100$, 200, 
and 1200. For carbon, these $r_{\rm s}$ correspond to 
mass densities $\rho \approx 1.6 \times 10^{13}$, $2 \times 10^{12}$, 
and $10^{10}$ g cm$^{-3}$, respectively. At such densities, carbon 
can hardly exist because it burns in nuclear reactions and suffers 
beta captures. Hence, practically important range for carbon seems to be 
$r_{\rm s} \gtrsim 1200$. For heavier elements the situation is similar. 
For helium, it is possible to attain lower values of $r_{\rm s}$. For 
instance, according to table 2 of \citet{PTY14}, helium may be found  
at $\rho \sim 10^8$ g cm$^{-3}$ ($r_{\rm s} \approx 140$) and 
$\Gamma \sim 1$, which, however, is not a very quantum state according 
to our Eq.\ (\ref{TTp}). Alternatively, in table 3 of the same work, 
helium exists at $\rho \sim 10^6$ g cm$^{-3}$ 
($r_{\rm s} \approx 650$) and $\Gamma \sim 50$, which is considerably 
more quantum.  

In the present paper, we have decided to focus on thermodynamics of 
the strongly-coupled liquid regime in the range of densities 
$r_{\rm s} \geq 600$, where reliable quantum results are largely absent. 
The case of higher densities at $\Gamma \sim 1$ is essentially classic 
(but may be worth a separate study).

\section{Path Integral Monte Carlo}
\label{PIMC}
Presently, only numerical techniques exist for first-principle 
studies of quantum strongly-coupled Coulomb plasma. This is true for 
both liquid and solid phases in which all anharmonic effects cannot be 
accounted for analytically. Following \citet{JC96}, we employ the PIMC 
method. Its detailed exposition is given in an extremely helpful review 
of \citet{C95}. 

In our simulation, there are $N=250$ distinguishable particles (ions) 
placed in a cubic simulation cell with periodic boundary conditions. 
Each particle interacts with all periodic images of the other particles 
and with the uniform background. Hence, any two particles interact not 
via Coulomb potential, but via Ewald potential, which is a sum of 
Coulomb potentials of all periodic images and the background term. In 
this way an approximate description of an infinite system is achieved.

In order to calculate various thermodynamic averages for a quantum 
system, it is represented by a classic one, in which $N$ quantum 
particles are replaced by $N$ classic ring polymers with 
$M$ numbered nodes (``beads'') each and $M$ links connecting beads with
consecutive numbers ($M$ is a positive integer). In each ring, there 
is an interaction between any two linked beads, which does not 
allow them to move arbitrarily far away from each other (``kinetic 
spring''). There is also an effective interaction between polymers 
originating from an interaction between beads having the 
same number in all rings.

If the effective interaction \citep[known as ``link action'', 
see][for details]{C95} were known exactly, then the classic and quantum 
systems would be exactly isomorphic for any $M$. However, this is 
not the case, and various approximations for the link action 
have been proposed. The simplest of them is known as the {\it primitive} 
approximation. Within its framework, the link action is directly 
related to the actual potential energy of $N$ beads having the same 
number in their respective rings and interacting only between 
themselves via the Ewald potential. The primitive approximation can be 
shown to become exact in the $M \to \infty$ limit.   

\citet{C95} emphasized the utility of another approximation, the 
{\it pair action}, for its ability to produce correct results at 
smaller $M$ than the primitive approximation in some problems 
\citep[see also,][]{MG06,M16}. A smaller $M$ translates into less 
time-consuming computations. We have managed to construct the pair 
action for the Ewald potential including off-diagonal terms for 
practically relevant ranges of action arguments (not too close to zero 
and not too far off the diagonal). However, it has been found that 
the primitive approximation reproduced the zero-point energy of 
a Coulomb crystal (the major contribution to which is given by the 
well-known harmonic lattice value) at smaller $M$ than the pair action. 
Consequently, we have chosen to use the primitive approximation for 
the action in this study. 

In the primitive approximation, one needs to sample the probability 
distribution 
\begin{equation}
     \pi(R_1, R_2, \ldots, R_M=R_0) = 
    \frac{1}{Z} \exp{\left[- \sum_{m=1}^M S_m  \right]}~,
\label{pi}
\end{equation}
\begin{equation}
         e^{-S_m} = \frac{1}{(4 \pi \lambda \tau)^{3N/2}}
      \exp{\left\{-\left[ \frac{(R_m-R_{m-1})^2}{4 \lambda \tau} 
      + \tau V(R_m) \right]\right\}},                             
\nonumber
\end{equation}
where $R_m$ denotes three Cartesian coordinates of all $N$ beads with 
number $m$, i.e., 750 numbers in our simulation (beads 0 are the 
same as beads $M$). Furthermore, $Z$ is the partition function, which 
ensures that the distribution $\pi$ is normalized, 
$\lambda = \hbar^2/(2M_i)$, $\tau = 1/(MT)$, and $V(R_m)$ is the 
potential energy of all beads number $m$. 

The sampling was done with the aid of the Metropolis algorithm. We 
have combined two types of moves: single bead moves using free-particle 
sampling and ``classic'' moves. In the latter we attempted to move 
a polymer as a whole using uniform sampling probability density within 
a cube whose size was chosen to ensure decent move acceptance rates. 
The latter moves have considerably improved the quality of our results 
at nearly classic $\rho$ and $T$.

The following quantity was used as the energy estimator (thermodynamic
estimator):
\begin{equation}
       E = \left\langle \frac{3N}{2 \tau} 
             - \frac{(R_m - R_{m-1})^2}{4 \lambda \tau^2} + V(R_m) 
               \right\rangle~,
\label{E_T}
\end{equation}
where $m$ is an arbitrary integer between 1 and $M$, to which the 
results should not be sensitive. Angle brackets mean averaging
with the probability distribution (\ref{pi}).

\section{Results}
\label{results}
\begin{figure*}                                           
%\vspace{-0.5cm}                                            
\begin{center}                                              
\leavevmode                                                 
\includegraphics[height=74mm,bb=72 529 539 740,clip]{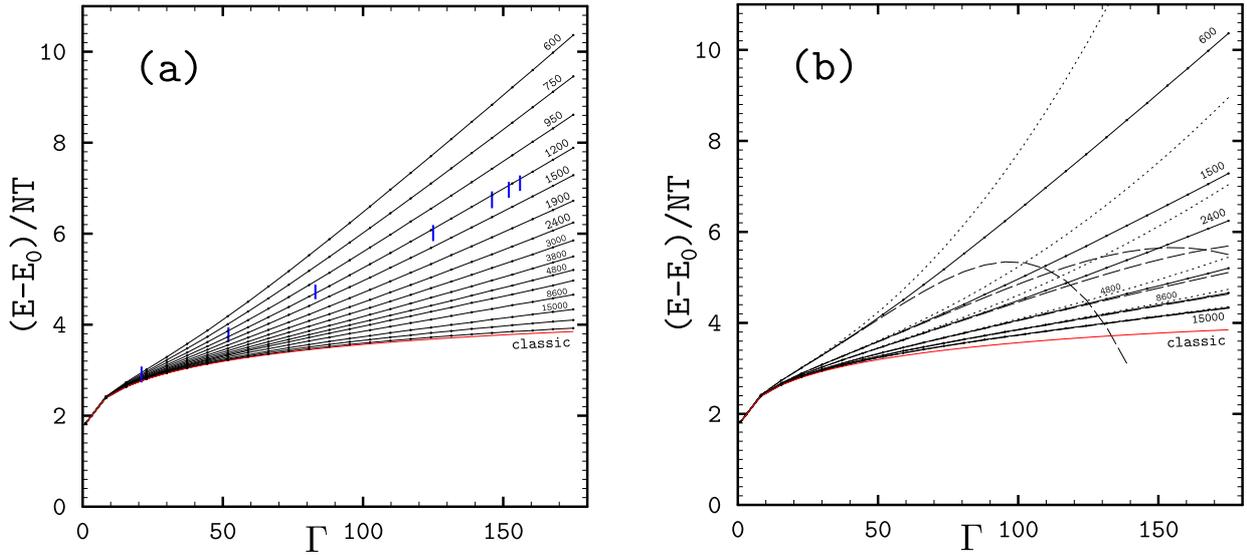} 
\end{center}                                                
\vspace{-0.4cm} 
\caption[]{(a) Calculated thermal energy of the Coulomb liquid 
vs $\Gamma$. From top to bottom: $r_{\rm s}=600$, 750, 950, 1200, 
1500, 1900, 2400, 3000, 3800, 4800, 6000, 8600, 15000, 30000, and 
120000. The curve on the very bottom (red) is 
the classic liquid energy. Bars represent selected data points 
of \citet{JC96} at $r_{\rm s}=1200$. (b) Same as in panel (a) but 
for $r_{\rm s}=600$, 1500, 2400, 4800, 8600, and 15000 only. 
Dotted and long-dashed curves incorporate, respectively, the first and 
the second quantum Wigner-Kirkwood corrections to the classic 
energy.}                                             
\label{figall}
\end{figure*}

Thermal energy of the liquid (equal to the total internal energy 
minus the Madelung energy of the body-centered cubic lattice, $E_0$), 
calculated by the method described in Sec.\ \ref{PIMC}, is presented in 
Fig.\ \ref{figall}(a) and Table \ref{enrg}. Dots 
in Fig.\ \ref{figall}(a) are computational results (including the 
extrapolation procedure below) while solid lines simply connect them to 
improve graph readability. We have calculated the 
thermal energy for 25 values of $\Gamma$ evenly spaced between 1 and 
175 ($\Delta \Gamma = 7.25$) and for 15 values of $r_{\rm s}=600$, 750, 
950, 1200, 1500, 1900, 2400, 3000, 3800, 4800, 6000, 8600, 15000, 
30000, 120000. This corresponds 
to $\sim 7$ orders of density variation and, e.g., for 
carbon, spans the range from $10^4$ to $8 \times 10^{10}$ g cm$^{-3}$ 
while for helium, the respective range of densities is from $0.2$ to 
$10^6$ g cm$^{-3}$. Calculations are unreliable at the lower end of 
densities and such $Z_i$ for which complete ionization is not reached, 
because electronic degrees of freedom are not taken into account.   
Solid curves with dots correspond to different $r_{\rm s}$ with the 
$r_{\rm s}=600$ curve being on top, $r_{\rm s}=750$ the second curve 
from the top, 
$r_{\rm s}=950$ the third, etc, and, finally, $r_{\rm s} = 120000$ is 
the second curve from the bottom. The (red) curve on the very bottom 
shows classic liquid thermal energy as calculated by \citet{C99} and 
fitted by \citet{PC00}.

%******************************************************************
%                                                       FIGURE 
%
%******************************************************************

\begin{figure*}                                           
%\vspace{-0.5cm}                                            
\begin{center}                                              
\leavevmode                                                 
\includegraphics[height=75mm,bb=71 526 539 741,clip]{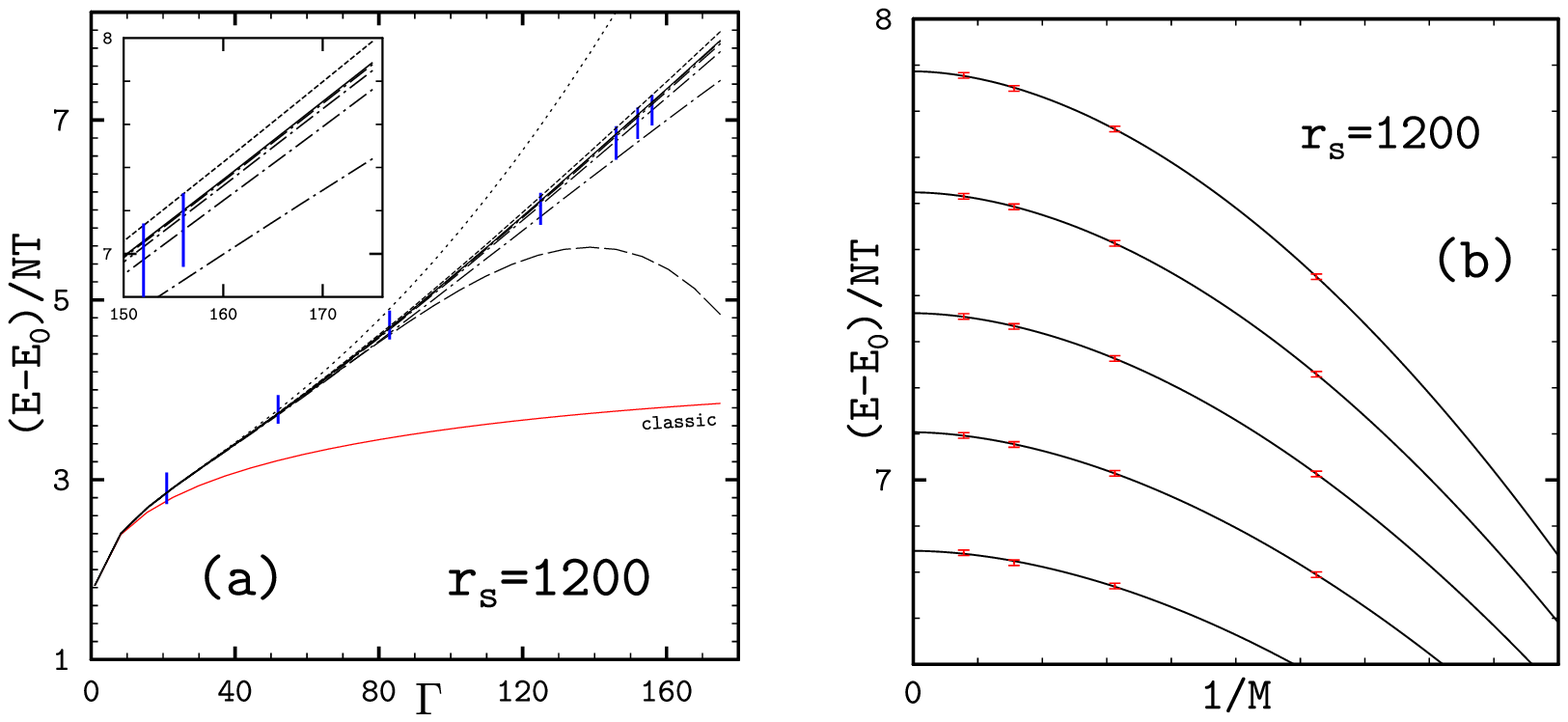} 
\end{center}                                                
\vspace{-0.4cm} 
\caption[]{(a) Various results at $r_{\rm s}=1200$. Four 
dot-dashed curves
are calculations at different $M$ (curves going higher correspond 
to bigger $M$). Solid curve (nearly merging with the top dot-dashed 
curve) shows the data extrapolated to 
$1/M=0$. Short-dashed curve includes the finite $N$ correction, 
Eq.\ (\ref{finN}). Dotted and long-dashed lines display the same 
Wigner-Kirkwood approximations as in Fig.\ \ref{figall}(b). 
Bars represent selected data points of \citet{JC96}.
(b) Extrapolation of calculated energies from finite $M$ to 
$1/M=0$ at $r_{\rm s}=1200$ and five biggest values of $\Gamma$.}                                             
\label{fig1200}
\end{figure*}

In Fig.\ \ref{figall}(b), we compare solid curves from 
Fig.\ \ref{figall}(a) for $r_s=600$, 1500, 2400, 4800, 8600, and 15000 
with the thermal energy of the classic liquid combined with the first 
quantum Wigner-Kirkwood correction \citep{H73} at the same $r_{\rm s}$  
(dotted curves). Long dashes also include  
the second quantum correction of \citet{HV75}. The agreement with our 
results is quite good in a more classic range of parameters but it gets
worse with increase of $r_{\rm s}$ or decrease of $T$ 
(i.e., increase of $\Gamma$). The first quantum correction provides an 
adequate description of our computed values for all $\Gamma$ 
in the liquid at $r_{\rm s} \gtrsim 20000$. The Hansen-Viellefosse
term extends this range down to $r_{\rm s} \gtrsim 10000$. 
[For $r_s=1200$, these approximations are also plotted in Fig.\ 
\ref{fig1200}(a).]    

\begin{table*}
\caption[]{Numerical results for $(E-E_0)/NT$}
\begin{tabular}{rrrrrrrrrrrrrrrr}
\hline      
\hline      
              & \multicolumn{15}{c}{$r_{\rm s}$} \\
                    \cline{2-16}
     $\Gamma$~~~~ & 600 & 750 & 950 & 1200 & 1500 & 1900 & 2400 & 3000 & 3800 & 4800 & 6000 & 8600 & 15e3 & 3e4 & 12e4 \\
\hline      
    175.00 &      10.364 &       9.457 &       8.613 &       7.886 &       7.285 &       6.723 &       6.244 &       5.851 &       5.499 &       5.202 &       4.970 &       4.659 &       4.336 &       4.102 &       3.924 \\
    167.75 &       9.979 &       9.118 &       8.313 &       7.624 &       7.050 &       6.519 &       6.070 &       5.700 &       5.367 &       5.086 &       4.862 &       4.581 &       4.281 &       4.067 &       3.901 \\
    160.50 &       9.595 &       8.778 &       8.020 &       7.362 &       6.820 &       6.320 &       5.893 &       5.544 &       5.234 &       4.975 &       4.763 &       4.500 &       4.226 &       4.028 &       3.871 \\
    153.25 &       9.216 &       8.440 &       7.723 &       7.103 &       6.591 &       6.120 &       5.719 &       5.396 &       5.101 &       4.860 &       4.665 &       4.430 &       4.169 &       3.989 &       3.850 \\
    146.00 &       8.832 &       8.103 &       7.426 &       6.846 &       6.363 &       5.920 &       5.546 &       5.247 &       4.973 &       4.753 &       4.570 &       4.346 &       4.115 &       3.946 &       3.818 \\
    138.75 &       8.457 &       7.771 &       7.131 &       6.588 &       6.138 &       5.723 &       5.379 &       5.095 &       4.844 &       4.640 &       4.480 &       4.273 &       4.052 &       3.908 &       3.788 \\
    131.50 &       8.080 &       7.436 &       6.838 &       6.332 &       5.913 &       5.530 &       5.210 &       4.948 &       4.718 &       4.526 &       4.379 &       4.193 &       3.997 &       3.861 &       3.756 \\
    124.25 &       7.705 &       7.103 &       6.549 &       6.077 &       5.686 &       5.332 &       5.038 &       4.804 &       4.594 &       4.418 &       4.286 &       4.113 &       3.937 &       3.820 &       3.723 \\
    117.00 &       7.335 &       6.775 &       6.267 &       5.826 &       5.468 &       5.143 &       4.873 &       4.655 &       4.468 &       4.310 &       4.188 &       4.038 &       3.882 &       3.769 &       3.685 \\
    109.75 &       6.968 &       6.449 &       5.977 &       5.575 &       5.252 &       4.954 &       4.712 &       4.513 &       4.343 &       4.205 &       4.098 &       3.958 &       3.821 &       3.723 &       3.648 \\
    102.50 &       6.599 &       6.127 &       5.698 &       5.331 &       5.035 &       4.767 &       4.546 &       4.371 &       4.221 &       4.097 &       3.998 &       3.880 &       3.756 &       3.672 &       3.608 \\
     95.25 &       6.235 &       5.807 &       5.418 &       5.088 &       4.827 &       4.584 &       4.388 &       4.236 &       4.096 &       3.992 &       3.904 &       3.804 &       3.696 &       3.622 &       3.568 \\
     88.00 &       5.878 &       5.492 &       5.141 &       4.848 &       4.623 &       4.405 &       4.232 &       4.098 &       3.979 &       3.885 &       3.812 &       3.721 &       3.633 &       3.568 &       3.521 \\
     80.75 &       5.526 &       5.184 &       4.872 &       4.618 &       4.411 &       4.224 &       4.083 &       3.963 &       3.862 &       3.779 &       3.717 &       3.643 &       3.562 &       3.510 &       3.470 \\
     73.50 &       5.176 &       4.876 &       4.607 &       4.389 &       4.215 &       4.053 &       3.930 &       3.830 &       3.740 &       3.675 &       3.622 &       3.556 &       3.494 &       3.448 &       3.416 \\
     66.25 &       4.837 &       4.578 &       4.352 &       4.165 &       4.018 &       3.885 &       3.779 &       3.698 &       3.624 &       3.569 &       3.526 &       3.471 &       3.419 &       3.384 &       3.357 \\
     59.00 &       4.504 &       4.289 &       4.100 &       3.944 &       3.825 &       3.716 &       3.626 &       3.563 &       3.507 &       3.460 &       3.428 &       3.387 &       3.340 &       3.313 &       3.296 \\
     51.75 &       4.186 &       4.010 &       3.859 &       3.731 &       3.636 &       3.551 &       3.479 &       3.427 &       3.384 &       3.351 &       3.325 &       3.290 &       3.257 &       3.236 &       3.220 \\
     44.50 &       3.871 &       3.735 &       3.619 &       3.524 &       3.450 &       3.386 &       3.334 &       3.296 &       3.263 &       3.234 &       3.215 &       3.190 &       3.168 &       3.152 &       3.141 \\
     37.25 &       3.578 &       3.477 &       3.387 &       3.321 &       3.267 &       3.221 &       3.186 &       3.159 &       3.132 &       3.114 &       3.100 &       3.084 &       3.065 &       3.054 &       3.048 \\
     30.00 &       3.290 &       3.225 &       3.167 &       3.117 &       3.083 &       3.051 &       3.029 &       3.009 &       2.995 &       2.982 &       2.973 &       2.962 &       2.951 &       2.942 &       2.937 \\
     22.75 &       3.014 &       2.969 &       2.940 &       2.911 &       2.892 &       2.873 &       2.858 &       2.847 &       2.839 &       2.832 &       2.828 &       2.820 &       2.814 &       2.809 &       2.808 \\
     15.50 &       2.735 &       2.715 &       2.697 &       2.686 &       2.677 &       2.668 &       2.660 &       2.657 &       2.652 &       2.648 &       2.645 &       2.644 &       2.640 &       2.638 &       2.636 \\
      8.25 &       2.418 &       2.414 &       2.408 &       2.405 &       2.404 &       2.401 &       2.397 &       2.395 &       2.395 &       2.396 &       2.393 &       2.394 &       2.392 &       2.391 &       2.392 \\
      1.00 &       1.825 &       1.825 &       1.825 &       1.825 &       1.826 &       1.826 &       1.825 &       1.825 &       1.825 &       1.825 &       1.826 &       1.826 &       1.825 &       1.825 &       1.825 \\
\hline      
\hline      
\end{tabular}
\label{enrg}
\end{table*}

There are three sorts of errors that need to be addressed. Firstly, 
there is a statistical error stemming from the limited number of 
steps in the Metropolis algorithm. This error manifests itself as a 
dependence of energies on initial conditions, random seeds, or the 
number $m$ in Eq.\ (\ref{E_T}). We have analysed these dependencies
for a few combinations of $r_{\rm s}$ and $T$ and concluded that the 
confidence interval could be estimated at the 90\% level as 
$\pm 0.007$ (in absolute units) for the entire dataset. This is 
smaller than the dot sizes in Fig.\ \ref{figall}(a). A much 
more extended computation is needed for a rigorous statistical 
evaluation of our data. 

Secondly, there is a systematic error associated with the finite 
polymer size $M$. In order to reduce this error, we have performed 
PIMC calculations with 4 values of $M$: $M_{\rm min}$, $2 M_{\rm min}$, 
$4 M_{\rm min}$, and $M_{\rm max}=8 M_{\rm min}$. For illustration, 
the respective energies at $r_{\rm s} = 1200$ are plotted in 
Fig.\ \ref{fig1200}(a) by dot-dashed curves: $M_{\rm max}$ is the top 
one and all the other go progressively lower with decrease of $M$. 
At each $r_{\rm s}$ and $\Gamma$, these data have been extrapolated  
as functions of $1/M$ down to $1/M = 0$. 
An example of this procedure is given in Fig.\ \ref{fig1200}(b) 
for $r_{\rm s} = 1200$ and several values of $\Gamma$. The resulting 
dependence represents the exact quantum energy of 250 distinguishable 
particles in periodic boundary conditions. For $r_{\rm s} = 1200$, it 
is plotted in Fig.\ \ref{fig1200}(a) by the solid line, nearly merging 
with the top dot-dashed line. It is these extrapolated data which are 
presented in Fig.\ \ref{figall}(a) and in Table \ref{enrg}. 

%******************************************************************
%                                                       FIGURE 
%
%******************************************************************

Bars in Figs.\ \ref{figall}(a) and \ref{fig1200}(a) represent the actual 
height and position of triangles from the bottom panel of fig.\ 3 of 
\citet{JC96}. We see, that at $r_{\rm s} = 1200$, our results are 
compatible with theirs.  

Thirdly, there is a systematic error due to the finite number of 
particles in our simulation, $N=250$. In principle, a conclusive study 
of $N$-dependence must include similar calculations at several 
different $N$ which is time-consuming and we defer it to a 
future work \citep[see also][]{Hetal16}. Based on studies of smaller 
systems, \citet{JC96} had come up with the following formula 
[their Eq.\ (2)]
\begin{equation}
      \frac{{\cal E}_N - {\cal E}_\infty}{T} 
      = [0.035(18) - 0.0018(1) \Gamma] \, \frac{\Gamma}{2N}~, 
\label{finN}
\end{equation}
where ${\cal E}_N$ and ${\cal E}_\infty$ are per-particle energies in 
the finite system and in the thermodynamic limit, respectively, 
and we have taken into account that their energies were expressed in 
ionic Rydbergs. This correction applied to $r_{\rm s}=1200$ produces, 
as the $N \to \infty$ limit, the short-dashed curve in 
Fig.\ \ref{fig1200}(a), which barely touches the bars. Remarkably, the 
correction does not depend on $r_{\rm s}$ [cf.\ the right-hand side of 
Eq.\ (\ref{finN})],
and thus it will be of about the same absolute magnitude at higher 
$r_{\rm s}$. But this can hardly be the case, because at high 
$r_{\rm s}$ our numerical data are in a much better agreement with 
classic results plus quantum corrections, which implies that 
at these $r_{\rm s}$, $N$-correction must be negligible. In view of 
this controversy, we have chosen not to apply the finite $N$ correction 
of \citet{JC96} in the present work pending a detailed future study.

\section{Discussion and outlook}
Our results can be used to derive thermodynamic properties of 
matter necessary for modelling various physical phenomena 
in compact stellar objects. 

First of all, the change of the ion energy due to quantum effects 
studied in the present paper, produces a change of the ion pressure and 
thus a slight modification of the equation of state in white 
dwarf cores and neutron star crusts. This effect is expected to be 
small though, because the total pressure is dominated by degenerate 
electrons.  

Thermal properties of matter are modified in a more meaningful way.
Consider, for instance, heat capacity. In a wide range of parameters 
the heat capacity is dominated by ions. This quantity is very important
for a number of astrophysical applications. For one it determines 
cooling of white dwarfs. Reliable modelling of this process allows one 
to interprete observed white dwarf luminosity function and extract such 
a fundamental parameter as the age of the local Galactic disk 
\citep[e.g.,][]{DM90}. 
Heat capacity of the Coulomb liquid is also needed to  
understand observed real-time cooling of neutron star crust heated 
during episodes of accretion in X-ray transients 
\citep[e.g.,][]{DCBR17}, although in this case 
temperatures may be too high for quantum effects to be 
noticeable.
Another promising research route is thermal evolution of accreting 
white dwarfs in cataclismic variables \citep[e.g.,][]{G00}.  

\begin{figure*}                                           
%\vspace{-0.5cm}                                            
\begin{center}                                              
\leavevmode                                                 
\includegraphics[height=77mm,bb=72 521 540 740,clip]{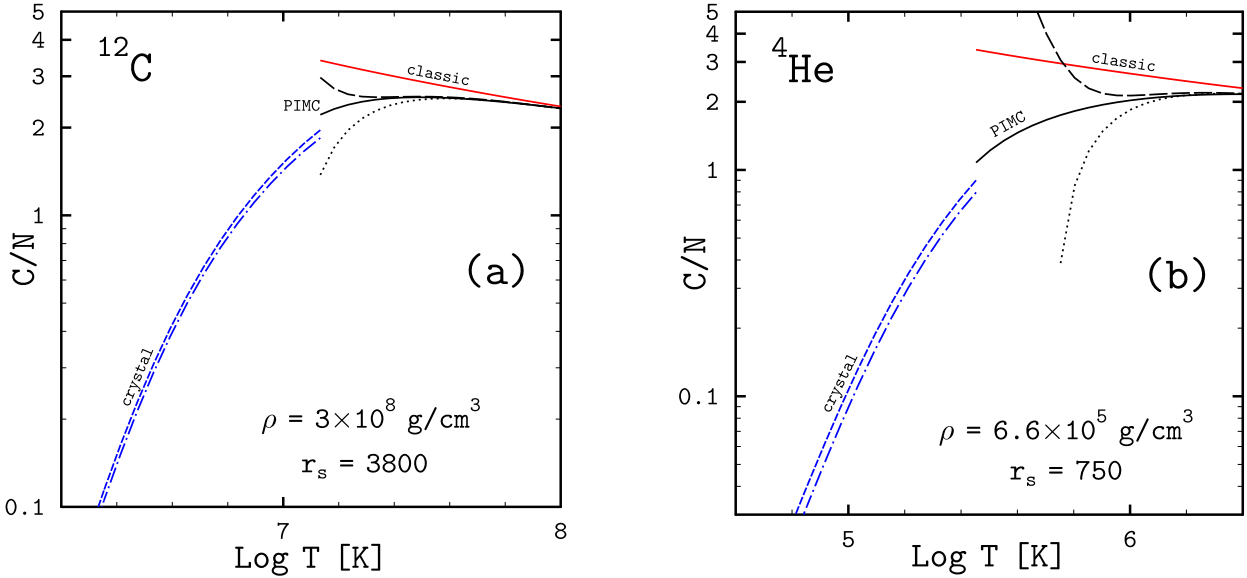} 
\end{center}                                                
\vspace{-0.4cm} 
\caption[]{(a) Ion specific heat of carbon at 
$\rho \approx 3 \times 10^8$ g cm$^{-3}$. Shown are PIMC results in 
the liquid phase (solid curve), specific heat of the classic liquid 
(upper solid curve, red), first quantum correction in the liquid 
(dots), second quantum correction in the liquid (long dashes), 
harmonic lattice results in the crystal phase (dot-dashed curve, 
blue), quantum anharmonic correction in the crystal (short dashes, 
blue). (b) Same as in panel (a) but for helium at 
$\rho \approx 6.6 \times 10^5$ g cm$^{-3}$.}
\label{carbon}
\end{figure*}

Let us analyse two representative examples:  
carbon at $\rho \approx 3 \times 10^8$ g cm$^{-3}$ 
($r_{\rm s} = 3800$) and helium at 
$\rho \approx 6.6 \times 10^5$ g cm$^{-3}$ ($r_{\rm s} = 750$). 
Ion specific heat at constant volume is 
\begin{equation}
    \frac{C}{N} = \frac{1}{N} \, 
    \left( \frac{\partial E}{\partial T} \right)_V~,
\label{CovN}
\end{equation}
where $E$ is the energy calculated in the previous section. 
The specific heat of carbon and helium is plotted in 
Fig.\ \ref{carbon}, panels (a) and (b), respectively, as a function of 
temperature accross the melting transition. For the sake of this 
illustration, we have assumed that 
melting takes place at $\Gamma_{\rm m}=175$. We are aware of the fact 
that $\Gamma_{\rm m}$ is a weak function of $r_{\rm s}$ due to quantum 
effects but an accurate determination of this function requires 
extra work and is beyond the scope of the present paper. 
At lower temperatures ($T$ to the left of the discontinuities), 
a body-centered cubic crystal phase takes place 
whose specific heat is known very accurately in the harmonic 
approximation \citep{BPY01}. It is shown by the (blue) dot-dashed 
curve. The (blue) short-dashed curve includes a quantum 
anharmonic contribution calculated in a model-dependent 
way by \citet{CB05}.

In the liquid phase, we plot various approximations 
mentioned in Sec.\ \ref{results}. Upper solid curve 
(red) shows the specific heat of the classic liquid OCP derived from 
fits of \citet{PC00}.  Dots include the 
first quantum correction to the classic values obtained from the 
Wigner-Kirkwood (WK) expansion \citep{H73}. 
The long-dashed curve includes the second WK correction derived 
by \citet{HV75}. Our PIMC results are 
shown by the solid curve. In order to plot it, we have fitted our energy 
data at fixed $r_{\rm s}$ by a smooth curve and differentiated
it with respect to $T$.

%******************************************************************
%                                                       FIGURE 
%
%******************************************************************

It is interesting to observe that the specific heat jump at melting  
(at quantum temperatures) appears to be much smaller with quantum 
effects included from first principles than with various approximations 
available previoulsly. A more complete assessment of this result
will be made when first-principle thermodynamics of the solid phase 
becomes available. More important, from practical point of view, is 
that our data indicate that the classic approximation for the 
specific heat can seriously overestimate it while an addition of the 
first WK correction yields an equally serious underestimation. 
The total heat capacity of a star, which determines its cooling rate, 
is an integral of the specific heat over the star. The degree to which 
the total heat capacity is affected by the quantum phenomena thus 
depends on the central density, temperature, and composition.  
A preliminary estimate suggests that in certain situations the actual 
heat capacity of a star may be as low as $2/3$ of the respective classic 
value.  

Besides carbon and helium, it is just as easy to consider heavier 
elements. However, in order to produce quantum effects of comparable 
size one would have to go to much higher densities. Since $r_{\rm s}$
scales as $\rho^{-1/3} Z_i^2 M_i^{4/3}$, to achieve $r_{\rm s}=3800$, 
one would require $\rho = 5.3 \times 10^9$ g cm$^{-3}$ for $^{16}$O and 
$\rho = 9.4 \times 10^{14}$ g cm$^{-3}$ for $^{56}$Fe (the latter is 
impossible because such $\rho$ exceeds the nuclear saturation density). 
In the standard picture, with growth of the mass density, 
lighter elements burn into heavier ones in nuclear reactions. The 
maximum density for a given light element to exist at a given 
temperature is a function of nuclear reaction rates in a 
strongly-coupled plasma. It is clear from the present study, that 
these reactions occur in the regime where quantum effects are well 
pronounced. Their rates are thus determined by the underlying quantum 
dynamics of ions and are extremely sensitive to small variations of 
density and temperature \citep[e.g.,][]{PM04}. Accordingly, it seems 
worthwhile to analyse the rates in detail from first principles which 
would require a minor modification of our code. Such a study would have 
implications for a wide range of astrophysical phenomena from bursts 
and flashes in compact stars to SN Ia.  

Another thermal property, which is expected to be sensitive to quantum 
effects is $\chi_T = ({\rm d} \ln{P}/{\rm d} \ln{T})_V$ where $P$ is 
the total pressure. This parameter is similar to the specific heat, 
Eq.\ (\ref{CovN}). In both cases, one needs to differentiate
a quantity dominated by degenerate electrons (energy in the case of $C$
and pressure in the case of $\chi_T$). However, the differentiation
is with respect to temperature which means that only thermal components
of $E$ and $P$ really matter. This leads to a domination of ion 
contributions over those due to degenerate electrons. 
Compressibility $\chi_T$ sets the scale of the Brunt-V{\"a}is{\"a}l{\"a} 
frequency \citep[e.g.,][]{BFWKT91}, which, in turn, determines main 
properties of stellar $g$-modes \citep[e.g.,][]{MW99}. Quantum effects 
may change the 
profile of the Brunt-V{\"a}is{\"a}l{\"a} frequency over a large 
fraction of the stellar radius. This may result in an observable change 
of the predicted seismological data for white dwarfs. 

The present work can be also extended in other directions.  
An accurate energy of the solid phase can be calculated which, among 
other things, would allow one to determine the dependence of the 
latent heat of crystallization on $r_{\rm s}$ due to quantum effects.
Latent heat of crystallization is another crucial parameter in the 
theory of white dwarf cooling. Its release delays cooling and directly 
affects the estimated age of a stellar population. 
A preliminary analysis (assuming fixed $\Gamma_{\rm m} = 175$) 
indicates that the latent heat is a weak function  
of $r_{\rm s}$ decreasing from $\approx 0.77\, T$ in the classic limit 
to $\approx 0.71\, T$ at $r_{\rm s}=1200$. A small correction to this 
dependence due to $r_{\rm s}$-dependence of $\Gamma_{\rm m}$  
is expected. Similar studies of multi-ionic mixtures can be 
performed with minor modifications of the same code. Finally, weak 
electron screening of the ion potential can be included but this will 
introduce an extra physical parameter into the problem.

\section*{Acknowledgements}
The author is sincerely greatful to A.\ A.\ Danilenko and 
D.\ G.\ Yakovlev for help and discussions. This work was 
supported by the Russian Science Foundation grant 19-12-00133.


\begin{thebibliography}{}

\bibitem[\protect\citeauthoryear{Baiko, Potekhin 
      \& Yakovlev}{2001}]{BPY01}
      Baiko D. A., Potekhin A. Y., and Yakovlev D. G., 2001, 
      Phys. Rev. E, 64, 057402 

\bibitem[\protect\citeauthoryear{Brassard et al.}{1991}]{BFWKT91}
      Brassard P., Fontaine G., Wesemael F., Kawaler S. D., and 
      Tassoul M., 1991, Astrophys. J., 367, 601 

\bibitem[\protect\citeauthoryear{Brush, Sahlin 
      \& Teller}{1966}]{BST66}
      Brush S. G., Sahlin H. L., and Teller E., 1966, 
      J. Chem. Phys., 45, 2102 

\bibitem[\protect\citeauthoryear{Caillol}{1999}]{C99}
      Caillol J. M., 1999, J. Chem. Phys., 111, 6538 

\bibitem[\protect\citeauthoryear{Ceperley}{1995}]{C95}
      Ceperley D. M., 1995, Rev. Mod. Phys., 67, 279

\bibitem[\protect\citeauthoryear{Chabrier, 
      Ashcroft \& DeWitt}{1992}]{CADW92}
      Chabrier G., Ashcroft N. W., and DeWitt H. E., 1992, 
      Nature, 360, 48

\bibitem[\protect\citeauthoryear{Chugunov \& Baiko}{2005}]{CB05}
      Chugunov A. I. and Baiko D. A., 2005, Physica A, 352, 397 

\bibitem[\protect\citeauthoryear{D'Antona \& Mazzitelli}{1990}]{DM90}
      D'Antona F. and Mazzitelli I., 1990, 
      Annu. Rev. Astron. Astrophys., 28, 139 

\bibitem[\protect\citeauthoryear{Deibel et al.}{2017}]{DCBR17}
      Deibel A., Cumming A., Brown E. F., and Reddy S., 
      2017, Astrophys. J., 839, 95  

\bibitem[\protect\citeauthoryear{DeWitt \& Slattery}{1999}]{DS99}
      DeWitt H. E. and Slattery W. L., 1999, 
      Contrib. Plasm. Phys., 39, 97 

\bibitem[\protect\citeauthoryear{G{\"a}nsicke}{2000}]{G00}
      G{\"a}nsicke B. T., 2000, Rev. Mod. Astron., 13, 151

\bibitem[\protect\citeauthoryear{Hansen}{1973}]{H73}
      Hansen J. P., 1973, Phys. Rev. A, 8, 3096 

\bibitem[\protect\citeauthoryear{Hansen \& Viellefosse}{1975}]{HV75}
      Hansen J. P. and Viellefosse P., 1975, Phys. Lett. A, 53, 187 

\bibitem[\protect\citeauthoryear{Holzmann et al.}{2016}]{Hetal16}
      Holzmann M., Clay R. C. III, Morales M. A., Tubman N. M., 
      Ceperley D. M., and Pierleoni C., 2016, Phys. Rev. B, 94, 035126

\bibitem[\protect\citeauthoryear{Jones \& Ceperley}{1996}]{JC96}
      Jones M. D. and Ceperley D. M., 1996, Phys. Rev. Lett., 76, 4572

\bibitem[\protect\citeauthoryear{Lamb \& Van Horn}{1975}]{LVH75}
      Lamb D. Q. and Van Horn H. M., 1975, Astrophys. J., 200, 306 

\bibitem[\protect\citeauthoryear{Militzer \& Graham}{2006}]{MG06}
      Militzer B. and Graham R. L., 2006, J. Phys. Chem. Sol., 67, 2136

\bibitem[\protect\citeauthoryear{Militzer}{2016}]{M16}
      Militzer B., 2016, Comp. Phys. Comm., 204, 88

\bibitem[\protect\citeauthoryear{Montgomery \& Winget}{1999}]{MW99}
      Montgomery M. H. and Winget D. E., 1999, Astrophys. J., 526, 976

\bibitem[\protect\citeauthoryear{Piersanti, Tornambe 
      \& Yungelson}{2014}]{PTY14}
      Piersanti L., Tornambe A., and Yungelson L. R., 2014, 
      Mon. Not. Roy. Astron. Soc., 445, 3239 

\bibitem[\protect\citeauthoryear{Pollock \& Militzer}{2004}]{PM04}
      Pollock E. L. and Militzer B., 2004, Phys. Rev. Lett., 92, 021101

\bibitem[\protect\citeauthoryear{Potekhin \& Chabrier}{2000}]{PC00}
      Potekhin A. Y. and Chabrier G., 2000, Phys. Rev. E, 62, 8554 

\bibitem[\protect\citeauthoryear{Slattery, Doolen 
      \& DeWitt}{1982}]{SDD82}
      Slattery W. L., Doolen G. D., and DeWitt H. E., 1982, 
      Phys. Rev. A, 26, 2255 

\bibitem[\protect\citeauthoryear{Van Horn}{1979}]{VH79}
      Van Horn H. M., 1979, Physics Today, Jan., 23  



\end{thebibliography}
\end{document}